\documentclass[11pt,a4paper]{article}

\usepackage[top=3cm, bottom=4cm, left=3cm, right=3cm]{geometry}

\usepackage{amsmath,amssymb}
\usepackage{amsthm}
\usepackage{accents}
\usepackage[active]{srcltx}
\usepackage{xspace}
\usepackage{cite}
\newcommand{\emi}{({\em i}\,)\xspace}
\newcommand{\emii}{({\em ii}\,)\xspace}

\newcommand{\R}{\mathbb{R}}

\newcommand{\la}{\lambda}
\newcommand{\eps}{\epsilon}
\newcommand{\veps}{\varepsilon}

\newcommand{\Ec}{{\cal E}}
\newcommand{\Lc}{{\cal L}}
\newcommand{\Mc}{{\cal M}}

\newcommand{\bld}[1]{\boldsymbol{#1}}
\newcommand{\we}{\wedge}
\newcommand{\ot}{\otimes}
\newcommand{\ovcirc}[1]{\accentset{\circ}{#1}}

\DeclareMathOperator{\hd}{\ast}
\DeclareMathOperator{\Div}{{\rm div}}

\DeclareMathOperator{\sgn}{{\rm sgn}}
\DeclareMathOperator{\lr}{\!\lrcorner}

\newcounter{mnotecount}[section]

\newtheorem{thr}{Theorem}

\numberwithin{equation}{section}
\numberwithin{thr}{section}

\begin{document}

%\title{Analysis of some solutions to constraints of the Teleparallel Equivalent of General Relativity}
\title{The Einstein constraints and differential forms}
\author{Andrzej Oko{\l}\'ow, Jakub Szymankiewicz}
\date{November 24, 2025}

\maketitle
\begin{center}
{\it  Institute of Theoretical Physics, University of Warsaw,\\ ul. Pasteura 5, 02-093 Warsaw, Poland\smallskip\\
oko@fuw.edu.pl}
\end{center}
\medskip

\begin{abstract}
We express the vacuum Einstein constraints in terms of differential forms---the forms include one-forms constituting an orthonormal coframe of the spatial metric. We show that if the metric is real-analytic, then the constraints can be always expressed locally as a system of first order PDE's---this system is obtained by a special choice of the coframe, which reduces to zero all second order terms in the scalar constraint.
\end{abstract}

%***************************************************
\section{Introduction}
%***************************************************

In the initial data formulation of general relativity, the admissible initial data is forced to satisfy so called Einstein constraints (see e.g. \cite{ein-constr-rev,ein-constr-rev2}), being four non-linear PDE's imposed on the data.

In the present paper we will prove that some simplified constraints \cite{tegr-con} on the phase space of the Teleparallel Equivalent of General Relativity (TEGR) are equivalent to the Einstein constraints in vacuum with zero cosmological constant. Thereby we will obtain a non-standard form of the Einstein constraints, which to the best of our knowledge has not been (extensively) studied so far.

The initial data in general relativity is usually encoded in \emi the spatial (Riemannian) metric induced on a Cauchy surface by the spacetime metric and \emii the extrinsic curvature of the surface. In that non-standard form of the Einstein constraints, the spatial metric is replaced by its orthonormal coframe $(\theta^I)$, $I\in\{1,2,3\}$, and the extrinsic curvature by a collection of two-forms $(r_J)$, $J\in\{1,2,3\}$, each being a momentum conjugate to a one-form constituting the coframe. In \cite{tegr-con} these constraints were expressed \emi in a component-less form, where the forms are combined by means of the exterior and interior products, the exterior derivative and the Hodge ``star'' given by the spatial metric and \emii in a component form, where additionally the two-forms $(d\theta^I)$ and $(r_J)$ were decomposed pointwisely into irreducible representations of the Lie group $SO(3)$.     

Based on the latter version of the constraints, a conjecture was formulated in \cite{tegr-con} that, given spatial metric, all second order terms in the constraints can be annihilated by a special choice of the orthonormal coframe. Applying a theorem by Bryant \cite{bry} we will show that locally the conjecture holds true for every real-analytic metric. This will allow us to rewrite the Einstein constraints as a system of first order PDE's imposed on the fields $(\theta^I)$ and $(r_J)$.    

The non-standard forms of the vacuum Einstein constraints considered in the present paper may prove fruitful for study of the initial value formulation of general relativity. In particular, they can be used to derive exact solutions to the constraints as will be illustrated with examples in the forthcoming paper \cite{prep-ao-js}.

The present paper is organised as follows. Section 2 contains preliminaries and a short summary of the relevant results obtained in \cite{tegr-con}. In Section 3 we will prove equivalence of the TEGR constraints and the vacuum Einstein ones. In Section 4 we will study the issue of the reduction of the Einstein constraints to a system of first order PDE's by means of a special choice of the coframe $(\theta^I)$. In Section 5 we will briefly describe other special orthonormal coframes found in the literature. The paper will conclude with Section 6 containing a brief summary of the results. 

%***************************************************
\section{Preliminaries}
%***************************************************

%***************************************************
\subsection{Phase space of TEGR}
%***************************************************

Let $\mathbb{E}$ be a three-dimensional real vector space equipped with a positive definite scalar product $\delta$. We fix an orthonormal basis $(e_1,e_2,e_3)\equiv (e_I)$, $I\in\{1,2,3\}$, of $\mathbb{E}$ and denote by $(\delta_{IJ})$ the components of $\delta$ in the basis. In the dual space $\mathbb{E}^*$ we choose a basis $(w^1,w^2,w^3)\equiv(w^I)$ dual to $(e_I)$. The scalar product $\delta$ induces on $\mathbb{E}^*$ the dual scalar product of components $(\delta^{IJ})$ in the basis $(w^I)$. By means of the components $(\delta_{IJ})$ and  $(\delta^{IJ})$ we will lower and raise indices $I,J,K,L,\ldots$.  

Let us also fix a three-dimensional oriented manifold $\Sigma$---the manifold will play the role of the standard leaf of a foliation being an element of $3+1$ decomposition of a spacetime. Consider now the following differential forms on $\Sigma$ : \emi $\xi$ and $\theta$, being, respectively, a zero-form (a function) and a one-form valued in $\mathbb{E}$ and \emii $r$ and $\zeta$, being, respectively, a two-form and a three-form valued in $\mathbb{E}^*$. The forms can be expressed in the bases $(e_I)$ and $(w^J)$:       
\begin{align*}
\xi&=\xi^I\ot e_I,& \theta&=\theta^J\ot e_J, & r&=r_K\ot w^K, & \zeta&=\zeta_L\ot w^L. 
\end{align*}

The fields above constitute the phase space of the canonical formulation of TEGR introduced in \cite{q-suit,ham-nv}: a collection $(\xi^I,\theta^J,r_K,\zeta_L)$ of real-valued differential forms on $\Sigma$ belongs to the phase space, provided the ordered set $(\theta^J)$ of one-forms is a global coframe on the manifold. The two-form $r_K$ is the momentum conjugate to the one-form $\theta^K$, and the three-form $\zeta_L$ is the momentum conjugate to the function $\xi^L$.    

%***************************************************
\subsection{The TEGR constraints}
%***************************************************

There are altogether ten constraints on the phase space of TEGR and all of them are of first class \cite{oko-tegr-II}. Six primary constraints generate gauge transformations on the phase space being an action of the Lorentz group \cite{oko-tegr-I,tegr-con}. There are four secondary constraints: three vector constraints and a scalar one. 

The primary result of \cite{tegr-con} is that all the functions $(\xi^I)$ can be always gauge-transformed to zero on the whole manifold $\Sigma$. In this gauge \emi the constraints become much simpler  and \emii the coframe $(\theta^I)$ becomes orthonormal with respect to the Riemannian metric $q$ induced on $\Sigma$ by the spacetime metric:   
\begin{equation}
q=\delta_{IJ}\,\theta^I\ot\theta^J.
\label{q}
\end{equation}
In the gauge $\xi^I=0$ the TEGR constraints read:
\begin{align}
&\veps^I\lr\zeta_I+\theta_J\we {\hd}d\theta^J=0,\label{bc}\\
&\veps^I\lr r_I=0,\label{rc}\\
& dr^J+d\theta^I\we \veps^J\lr r_I=0,\label{vc}\\
&[{\hd}(r_I\we\theta^J)][{\hd}(r_J\we\theta^I)]-\frac{1}{2}[{\hd}(r_I\we\theta^I)]^2+\nonumber\\+&2{\hd}d(\theta_J\we {\hd}d\theta^J)+[{\hd}(d\theta_I\we\theta^J)][{\hd}(d\theta_J\we\theta^I)]-\frac{1}{2}[{\hd}(d\theta_I\we\theta^I)]^2=0.
\label{*scal}
\end{align}
In the formulas above $(\veps_I)$ is the frame dual to $(\theta^J)$, and $*$ denotes the Hodge dualization (``star'') defined by the metric \eqref{q} and the orientation of $\Sigma$. The constraints \eqref{bc} and \eqref{rc} were called in \cite{oko-tegr-I} {\em boost} and {\em rotation} constraint respectively---they constitute the set of primary constraints of TEGR. The constraint \eqref{vc} is the vector constraint, while \eqref{*scal} is the scalar constraint.   

The fact that in the gauge $\xi^I=0$ the coframe $(\theta^J)$ is orthonormal w.r.t. the metric $q$, motivated the author of \cite{tegr-con} \emi to describe tensor fields occuring in the constraints by means of their components in $(\theta^I)$ and \emii to decompose {\em pointwisely} the fields $(r^I)$ and $(d\theta^I)$ into irreducible representations of the $SO(3)$ group preserving the scalar product $\delta$ on $\mathbb{E}$. In the resulting expressions, except the components of the momenta $(r^I)$ and $(\zeta_J)$, there appear components of the derivatives $(d\theta^I)$ and those of the volume form $\eps$ on $\Sigma$ defined by the metric $q$ and the orientation of the manifold:               
\begin{align}
r^I&=r^I{}_{JK}\,\theta^J\ot\theta^K, & d\theta^I&=\beta^I{}_{JK}\,\theta^J\ot\theta^K, \label{dth}\\
\zeta_{I}&=\zeta_{IJKL}\, \theta^J\ot\theta^K\ot\theta^L, & \eps&=\eps_{JKL}\,\theta^J\ot\theta^K\ot\theta^L. \nonumber 
\end{align}
The decomposition of $(r^I)$ and $(d\theta^J)$ reads:
\begin{align}
r^I{}_{JK}&=\frac{1}{2}r^{IL}\eps_{LJK}, & r^{IL}&:=r^{I}{}_{JK}\eps^{JKL}=r^{[IL]}+\ovcirc{r}^{IL}+\frac{1}{3}r^{K}{}_K\delta^{IL}, \label{r-dec} \\
\beta^I{}_{JK}&=\frac{1}{2}\beta^{IL}\eps_{LJK}, & \beta^{IL}&:=\beta^{I}{}_{JK}\eps^{JKL}=\beta^{[IL]}+\ovcirc{\beta}^{IL}+\frac{1}{3}\beta^{K}{}_K\delta^{IL},\label{b-dec}\\
\beta^{[IL]}&=\eps^{ILJ}\beta_J, & \beta_{I}&:=\frac{1}{2}\eps_{IJK}\beta^{JK},\label{b-dec'} 
\end{align}
where $(\ovcirc{r}^{I}{}_L)$ and $(\ovcirc{\beta}^{I}{}_L)$ are  fields on $\Sigma$, which are valued in traceless symmetric $3\times 3$ matrices. The fields $(\beta^{[IJ]})$ and $(\beta_I)$ will be called {\em antisymmetric} part of the derivatives $(d\theta^I)$, $(\ovcirc{\beta}^{IJ})$ {\em traceless symmetric} part and finally $\beta^{K}{}_K$ will be called {\em trace} of $(d\theta^I)$. Similar terminology will be used w.r.t. $(r^{[IJ]})$, $(\ovcirc{r}^{IJ})$ and $r^K{}_K$. Let us also note that
\begin{align}
2*r^I&=r^{I}{}_J\,\theta^J, & 2*d\theta^I&=\beta^{I}{}_J\,\theta^J.
\label{*r*th}
\end{align}

In the gauge $\xi^I=0$ the boost and rotation constraints, when expressed in terms of the decomposition above, take the following form:     
\begin{align}
&\zeta_{IJKL}+\frac{1}{4}\beta_I\eps_{JKL}=0, & & r^{[IJ]}=0.
\label{pri-con}
\end{align}
Clearly, this simple formulas can be treated as general solutions to the primary constraints: the momentum $\zeta_I$ is fully determined as a function of the coframe $(\theta^I)$, and the antisymmetric part of $(r^I)$ must vanish. In the same gauge the vector and the scalar constraints read, respectively,
\begin{align}
&\frac{2}{3}r^K{}_K{}^{,J}+2\ovcirc{r}^{JM}{}_{,M}+3\ovcirc{r}^{JM}\beta_M+\ovcirc{r}^{IM}\ovcirc{\beta}_{IN}\eps^{JN}{}_M=0,\label{vec}\\
& \ovcirc{r}^{IJ}\ovcirc{r}_{IJ}-\frac{1}{6}(r^K{}_K)^2+8\beta^I{}_{,I}+6\beta^I\beta_I+\ovcirc{\beta}^{IJ}\ovcirc{\beta}_{IJ}-\frac{1}{6}(\beta^K{}_K)^2=0. \label{scal}
\end{align}
In both expressions above we used the following notation:
\begin{align*}
f_{,I}&\equiv\varepsilon_If, & f^{,I}&\equiv\delta^{IJ}f_{,J}=\varepsilon^If.
\end{align*}
where $f$ is a function on $\Sigma$, and $(\varepsilon_I)$ is the frame dual to $(\theta^I)$.

Note now that to solve all the TEGR constraints in the gauge $\xi^I=0$ it is sufficient to solve the rotation constraint being an algebraic condition and the vector and scalar constraints being PDE's imposed on the coframe $(\theta^J)$ and the momenta $(r_K)$. Once we have a solution to the three constraints we can easily solve the boost constraints for the momenta $(\zeta_L)$. Therefore the boost constraint will be neglected in our considerations below.  

%***************************************************
\section{The TEGR constraints versus the vacuum Einstein constraints}
%***************************************************

Since in the case of zero cosmological constant TEGR is dynamically equivalent to vacuum general relativity, one can expect that the TEGR constraints are equivalent to the vacuum Einstein constraints. Below we will show by direct calculation that the rotation, vector and scalar constraints of TEGR in the gauge $\xi^I=0$ are equivalent to the Einstein constraints.

In the calculations below the following well-known formulas will be used repeatedly: 
\begin{align}
\eps^{IJK}\eps_{LMK}&=2\,\delta^{[I}{}_L\delta^{J]}{}_M, & \eps^{IJK}\eps_{LJK}&=2\,\delta^{I}{}_L, & {\hd}(\theta^I\we\theta^J\we\theta^K)&=\eps^{IJK}.
\label{var-eps}
\end{align}
Moreover, we will also refer to equations
\begin{align}
d\theta_I+\Gamma_{IJ}\we\theta^J&=0, & \Gamma_{IJ}&=-\Gamma_{JI},
\label{Gamma}
\end{align}
which are satisfied by the Levi-Civita connections one-forms  $(\Gamma_{IJ})$ given by the metric \eqref{q} and expressed in its orthonormal coframe $(\theta^I)$.

Suppose now that a coframe $(\theta^I)$ and momenta $(r_I)$ satisfy the three TEGR constraints. Then the first equation in \eqref{Gamma} allows us to rewrite the second term at the l.h.s. of the vector constraint \eqref{vc} as follows:
\begin{multline*}
d\theta^I\we\veps^J\lr r_I=-\Gamma^I{}_{K}\we\theta^K\we\veps^J\lr r_I = \veps^J\lr(\Gamma^I{}_{KL}\theta^L\we\theta^K)\we r_I=\Gamma^I{}_K{}^J\theta^K\we r_I-\Gamma^{IJ}{}_{L}\theta^L\we r_I
\end{multline*}
(here we used the components $(\Gamma^I{}_{KL})$ of $(\Gamma^I{}_K)$ in the coframe $(\theta^J)$). The first of the two resulting terms vanishes. To show this let us act on it by the Hodge ``star'' and express $r_I$ in terms of the components $(r_{IJ})$ (see the first equations in \eqref{dth} and \eqref{r-dec}),
\begin{equation}
r_I=\frac{1}{4}r_I{}^L\eps_{LMN}\,\theta^M\we\theta^N.
\label{rI-rIJ}
\end{equation}
This gives us what follows:
\begin{multline*}
*(\Gamma^I{}_K{}^J\theta^K\we r_I)=*(\Gamma^I{}_K{}^J\theta^K\we\frac{1}{4}r_I{}^L\eps_{LMN}\,\theta^M\we\theta^N)=\frac{1}{4}\Gamma^I{}_K{}^Jr_I{}^L\eps_{LMN}\eps^{KMN}=\\=\frac{1}{2}\Gamma^I{}_K{}^Jr_I{}^K=\frac{1}{2}\Gamma_{IK}{}^Jr^{IK}=0
\end{multline*}
---the last equality holds by virtue of the second equation in \eqref{Gamma} and the lack of the antisymmetric part in $(r^{IK})$ implied by the rotation constraint. Thus
\[
d\theta^I\we\veps^J\lr r_I=-\Gamma^{IJ}{}_{L}\,\theta^L\we r_I
\]
and the vector constraint takes thereby the following simple form:
\[
dr_J-\Gamma^I{}_J\we r_I=0.
\]

Applying again Equation \eqref{rI-rIJ} and the first equation in \eqref{Gamma} we transform the constraint further:
\begin{multline*}
0=d(r_{JL}\eps^L{}_{MN}\,\theta^M\we\theta^N)-\Gamma^{I}{}_{JK}r_{IL}\eps^L{}_{MN}\,\theta^K\we\theta^M\we\theta^N=\\=r_{JL,K}\eps^L{}_{MN}\,\theta^K\we\theta^M\we\theta^N-2r_{JL}\eps^L{}_{MN}\,\Gamma^M{}_{IK}\,\theta^K\we\theta^I\we\theta^N-\Gamma^{I}{}_{JK}r_{IL}\eps^L{}_{MN}\,\theta^K\we\theta^M\we\theta^N.
\end{multline*}
Let us now apply the Hodge ``star'' to both sides of the equation above:
\begin{multline}
0=r_{JL,K}\eps^L{}_{MN}\eps^{KMN}-2r_{JL}\eps^L{}_{MN}\,\Gamma^M{}_{IK}\eps^{KIN}-\Gamma^{I}{}_{JK}r_{IL}\eps^L{}_{MN}\eps^{KMN}=\\=2(r_{JL}{}^{,L}-\Gamma^M{}_M{}^Lr_{JL}-\Gamma^{LK}{}_Kr_{JL}-\Gamma^{I}{}_{J}{}^Lr_{IL})=2r_{JL}{}^{;L}
\label{divr=0}
\end{multline}
---here in the last step we used the second equation in \eqref{Gamma} to zero the contraction $\Gamma^M{}_M{}^L$. The semicolon in the result above denotes obviously the Levi-Civita covariant derivative given by the metric $q$. 

It is shown in Appendix \ref{r-K-app} that if $\xi^I=0$, then the momenta $(r_I)$ obtained by a suitable Legendre transformation (and a subsequent canonical transformation), are related to the extrinsic curvature\footnote{Let us recall that the extrinsic curvature $K$ is symmetric: $K_{IJ}=K_{JI}$.} $K=K_{IJ}\,\theta^I\ot\theta^J$ of $\Sigma$ as follows\footnote{In Equations \eqref{r-K} the components $(r_{IJ})$ are symmetric: $r_{IJ}=r_{JI}$, which means that the two-forms $(r_I)$ satisfy the rotation constraints $r_{[IJ]}=0$. This is because this constraint is a primary one and canonical variables belonging to the image of the Legendre transformation have to satisfy all the primary constraints.}:
\begin{align}
r_{IJ}&=2(K_{IJ}-K^L{}_L\delta_{IJ}), & K_{IJ}&=\frac{1}{2}r_{IJ}-\frac{1}{4}r^K{}_K\delta_{IJ}, \label{r-K}\\
r^L{}_L&=-4 K^L{}_L, & \ovcirc{r}_{IJ}&=2\ovcirc{K}_{IJ},\label{r-K-dec}
\end{align}
where $\ovcirc{K}_{IJ}:=K_{IJ}-K^L{}_L\,\delta_{IJ}/3$ is the traceless part of $K$.  

Now, to obtain the standard vector constraint (expressed in an orthonormal coframe),
\begin{equation}
(K_{IJ}-K^L{}_L\delta_{IJ})^{;J}=0,
\label{vec-std}
\end{equation}
it is enough to replace in \eqref{divr=0} the variables $(r_{IJ})$ by $(K_{IJ})$ according to the first equation in \eqref{r-K}.

Regarding the scalar constraint, if $(\theta^I)$ is an orthonormal coframe of a Riemannian metric $q$, then the Ricci scalar $\rho$ of $q$ can be written as a function of the irreducible parts of $(d\theta^I)$:
\begin{equation}
\rho=-\frac{1}{4}\Big(8\beta^K{}_{,K}+6\beta^K\beta_K+\ovcirc{\beta}_{IK}\ovcirc{\beta}^{IK}-\frac{1}{6}(\beta^K{}_K)^2\Big)
\label{R-sc}
\end{equation}
(for a derivation of this formula see Appendix \ref{ricci-app}). Applying this equation as well as the formulas \eqref{r-K-dec} to the scalar constraint \eqref{scal} one easily obtains the following equation
\begin{equation}
\rho-K_{IJ}K^{IJ}+(K^L{}_L)^2=0
\label{scal-std}
\end{equation}
being the standard vacuum scalar constraint.

The conclusion is that the rotation, vector and scalar constraints of TEGR in the gauge $\xi^I=0$ imply the vacuum Einstein constraints. 

On the other hand, let us assume that a metric $q$ and an extrinsic curvature $K$ on $\Sigma$ satisfy the vacuum Einstein constraints. Let us choose an orthonormal coframe $(\theta^I)$ of the metric $q$ and define the corresponding momenta $(r_I)$ by combining Equation \eqref{rI-rIJ} and the first formula in \eqref{r-K}. Note that the latter formula guarantees that $r_{[IJ]}=0$, which means that the resulting $(r_K)$ satisfy the rotation constraint. Then starting from the constraints \eqref{vec-std} and \eqref{scal-std} one can arrive at the vector and scalar constraints of TEGR by performing in reverse order the calculations presented above.       

Thus the rotation, vector and scalar constraints of TEGR in the gauge $\xi^I=0$ are equivalent to the vacuum Einstein constraints. Therefore the former constraints can be treated as an equivalent form of the latter ones. 

%***************************************************
\section{The Einstein constraints as a system of first order PDE's}
%***************************************************

The paper \cite{tegr-con} considered a possibility of further simplifying the TEGR constraints. It was noted there that in the constraints in the gauge $\xi^I=0$, there is only one term, which contains second derivatives, namely the term $\beta^I{}_{,I}$ in the scalar constraint \eqref{scal}. Consequently, if the derivatives $(d\theta^I)$ of a coframe lack their antisymmetric part, $\beta_I=0$, then the scalar constraint reduces to a first order PDE. A conjecture was formulated that every Riemannian metric $q$ on $\Sigma$ admits an orthonormal coframe $(\theta^J)$ such that the antisymmetric part of $(d\theta^I)$ vanishes. However, in that paper the conjecture was left without any proof.

Below we will show that the conjecture is locally true for real-analytic metrics. Then we will express the constraints as a system of first order PDE's.

%***************************************************
\subsection{Antisymmetric part of $(d\theta^I)$ }
%***************************************************

The antisymmetric part of $(d\theta^I)$ is related to various quantities, as shown in the following formulas, where either $\beta_I$ or $\beta^{[IJ]}$ appears:    
\begin{align}
\beta^J{}_{IJ}&=\beta_I, & \theta_I\we*d\theta^I&=\frac{1}{2}\beta^{[IJ]}\theta_I\we\theta_J, \label{bI-bJIJ}\\
\theta^J([\veps_I,\veps_J])&=-\beta_I, & \Lc_{\veps_I}\Ec&=-\beta_I\,\Ec,\label{Le}\\
\Lc_{\veps_I}\eps&=\beta_I\,\eps, & \Div_q\veps_I&=\beta_I,\label{div}\\
*d(\theta^I\we\theta^J)&=\beta^{[IJ]}, & *\,d*\theta^I&=\beta^I\label{coclsd},\\
\Gamma^J{}_{IJ}&=\beta_I. && \label{G-beta}
\end{align}
Let us now explain the meaning of some symbols used in these equations. $\Lc_{\veps_I}$ denotes the Lie derivative w.r.t. the vector field $\veps_I$ being an element of the frame dual to $(\theta^J)$ (i.e. $\theta^J(\veps_I)=\delta^J{}_I$). If $X$ is a vector field, then $\Div_qX$ is its divergence given by the Levi-Civita covariant derivative derived from the metric \eqref{q}---in a local coordinate system $(x^i)$ on $\Sigma$ $\Div_qX=X^i{}_{;i}$. The symbol $\Ec$ denotes a totally antisymmetric tensor field of type $\binom{3}{0}$ defined as the exterior product\footnote{Suppose that $\tau$ and $\bar{\tau}$ are totally antisymmetric tensor fields of type, respectively $\binom{k}{0}$ and $\binom{\bar{k}}{0}$. Then
\[
\tau\we \tau':=\frac{(k+\bar{k})!}{k!\bar{k}!} {\cal A}(\tau\ot\bar{\tau}),
\]
where ${\cal A}(\tau\ot\bar{\tau})$ denotes the antisymmetrization of the tensor field $\tau\ot\bar{\tau}$ w.r.t. all its arguments/indices.} of the vector fields $(\veps_I)$: 
\[
\Ec:=\veps_1\we\veps_2\we\veps_3.
\]
Finally, $(\Gamma_{IJK})$ are components of the connection one-forms $(\Gamma_{IJ})$ (see \eqref{Gamma}) in the coframe $(\theta^K)$. 

The left equation in \eqref{bI-bJIJ} is a simple consequence of the definition of $(\beta_I)$ (the second equation in \eqref{b-dec'}) and the definition of $(\beta^{IJ})$ (the second equation in \eqref{b-dec}). Note that this equation allows to express $(\beta_I)$ in a very simple way without any reference to the metric $q$ given by $(\theta^J)$.  

The right equation in \eqref{bI-bJIJ} was proven in \cite{tegr-con}.

It follows from \eqref{dth} that   
\begin{equation}
\beta^I{}_{JK}=(d\theta^I)(\veps_J,\veps_K).
\label{b-dth}
\end{equation}
On the other hand it is well-known that for any one-form $\omega$ and any vector fields $X$ and $Y$  
\begin{equation*}
(d\omega)(X,Y)=X(\omega(Y))-Y(\omega(X))-\omega([X,Y]).
\end{equation*}
These two formulas applied in turn to the left equation in \eqref{bI-bJIJ} gives us the left equation in \eqref{Le}.

To justify the right equation in \eqref{Le} let us calculate the Lie derivative\footnote{For any totally antisymmetric tensor fields $\tau$ and $\bar{\tau}$ of type, respectively, $\binom{k}{0}$ and $\binom{\bar{k}}{0}$ and any vector field $X$ the standard Leibniz rule is satisfied:  $\Lc_X(\tau\we\bar{\tau})=(\Lc_X\tau)\we\bar{\tau}+\tau\we\Lc_X\bar{\tau}$.} of the field $\Ec$ w.r.t. $\veps_I$:    
\begin{equation*}
\Lc_{\veps_I}\Ec=[\veps_I,\veps_1]\we\veps_2\we\veps_3+\veps_1\we[\veps_I,\veps_2]\we\veps_3+\veps_1\we\veps_2\we[\veps_I,\veps_3].
\end{equation*}
Since 
\[
[\veps_I,\veps_L]=\theta^K([\veps_I,\veps_L])\veps_K,
\]
the Lie derivative can be expressed as follows:
\begin{multline*}
\Lc_{\veps_I}\Ec=\theta^1([\veps_I,\veps_1])\veps_1\we\veps_2\we\veps_3+\veps_1\we\theta^2([\veps_I,\veps_2])\veps_2\we\veps_3+\veps_1\we\veps_2\we\theta^3([\veps_I,\veps_3])\veps_3=\\=\theta^J([\veps_I,\veps_J])\,\Ec=-\beta_I\,\Ec,
\end{multline*}
where in the last step we used the left equation in \eqref{Le}.

Similarly,
\begin{equation}
\Lc_{\veps_I}(\theta^1\we\theta^2\we\theta^3)=(\Lc_{\veps_I}\theta^1)\we\theta^2\we\theta^3+\theta^1\we(\Lc_{\veps_I}\theta^2)\we\theta^3+\theta^1\we\theta^2\we(\Lc_{\veps_I}\theta^3).
\label{L-eps}
\end{equation}
The Lie derivative of each one-form $\theta^J$ reads 
\[
\Lc_{\veps_I}\theta^J=\veps_I\lr d\theta^J+d(\veps_I\lr\theta^J)=\veps_I\lr d\theta^J,
\]
since $\veps_I\lr\theta^J=\delta^J{}_I$ is constant. On the other hand, every one-form $\omega$ is equal to $\omega(\veps_K)\,\theta^K$. Therefore   
\[
\Lc_{\veps_I}\theta^J=(\veps_I\lr d\theta^J)(\veps_K)\,\theta^K=d\theta^J(\veps_I,\veps_K)\,\theta^K=\beta^J{}_{IK}\,\theta^K,
\]
where in the last step we used \eqref{b-dth}. Inserting this result to \eqref{L-eps} we obtain
\[
\Lc_{\veps_I}(\theta^1\we\theta^2\we\theta^3)=\beta^L{}_{IL}\,\theta^1\we\theta^2\we\theta^3=\beta_I\,\theta^1\we\theta^2\we\theta^3.
\]
Clearly, the three-form $\theta^1\we\theta^2\we\theta^3$ is equal modulo the factor $\pm 1$ to the volume form $\eps$ of the metric $q$, so the left equation in \eqref{div} follows.  

It is well-known that for every vector field $X$ 
\[
\Lc_X\eps=(\Div_qX)\eps.
\]
Comparing this equation with the left one in \eqref{div} we obtain the right equation in \eqref{div}.

Using \eqref{dth} and \eqref{b-dec} we calculate:
\begin{equation}
*(d\theta^I\we\theta^J)=*\Big(\frac{1}{2}\beta^I{}_{MN}\,\theta^M\we\theta^N\we\theta^J\Big)=\frac{1}{2}\beta^I{}_{MN}\eps^{MNJ}=\frac{1}{2}\beta^{IJ}.
\label{ttIJ-bIJ}
\end{equation}
Consequently,
\[
*\big(d(\theta^I\we\theta^J)\big)=*(d\theta^I\we\theta^J-\theta^I\we d\theta^J)=*(d\theta^I\we\theta^J-d\theta^J\we \theta^I)=\frac{1}{2}(\beta^{IJ}-\beta^{JI})=\beta^{[IJ]}. 
\]
Thus we obtain the left equation in \eqref{coclsd}.

Contracting both sides of the equation above with $\eps^L{}_{IJ}/2$ and taking into account that 
\[
\frac{1}{2}\eps^L{}_{IJ}\,\theta^I\we\theta^J=*\theta^L
\]
we obtain the right equation in \eqref{coclsd}.

Finally, Equation \eqref{G-beta} is a direct consequence of Equation \eqref{G-beta-dec}.

Based on the equations \eqref{bI-bJIJ}--\eqref{G-beta} one sees that the derivatives $(d\theta^J)$ lack their antisymmetric part if and only if any of the following conditions holds:  
\begin{align}
\beta^J{}_{IJ}&=0, & \theta_I\we*d\theta^I&=0, \nonumber\\
\theta^J([\veps_I,\veps_J])&=0, & \Lc_{\veps_I}\Ec&=0,\label{eIE}\\
\Lc_{\veps_I}\eps&=0, & \Div_q\veps_I&=0,\label{div-q}\\
d(\theta^I\we\theta^J)&=0, & *\,d*\theta^I&=0,\label{coclsd-1}\\
\Gamma^J{}_{IJ}&=0. &&
\end{align}

%***************************************************
\subsection{Metrics admitting coclosed orthonormal coframes  }
%***************************************************

Let $q$ be a Riemannian metric on $\Sigma$. We will say that its orthonormal coframe $(\theta^J)$ is {\em coclosed}, if  
\[
*\,d*\theta^J=0.
\]
By virtue of the right equation in \eqref{coclsd-1} we can state that the antisymmetric part of the derivatives $(d\theta^J)$ vanishes, $\beta_I=0$, if and only if the coframe $(\theta^J)$ is coclosed.

The following theorem was proven by Bryant in \cite{bry} (see Remark 3 therein):
\begin{thr}
Let $q$ be a real-analytic Riemannian metric defined on a three-dimen\-sio\-nal manifold. Then there exists locally an orthonormal coframe of the metric that is coclosed.
\label{bry-thr}
\end{thr}

%***************************************************
\subsection{The system of first order PDE's}
% ***************************************************

Suppose that a real-analytic metric $q$ and an extrinsic curvature $K$ on $\Sigma$ satisfy the vacuum Einstein constraints. Let $(\theta^J)$ be a local coclosed coframe on $\Sigma$, which is orthonormal w.r.t. $q$---its existence is guaranteed by Theorem \ref{bry-thr}. Then $(\theta^J)$ and the corresponding momenta $(r_I)$ satisfy the following equations:         
\begin{align}
\beta_I&=0, & \frac{2}{3}r^K{}_K{}^{,J}+2\ovcirc{r}^{JM}{}_{,M}+\ovcirc{r}^{IM}\ovcirc{\beta}_{IN}\eps^{JN}{}_M&=0,\label{vec-f}\\
r^{[IJ]}&=0, &  \ovcirc{r}^{IJ}\ovcirc{r}_{IJ}-\frac{1}{6}(r^K{}_K)^2+\ovcirc{\beta}^{IJ}\ovcirc{\beta}_{IJ}-\frac{1}{6}(\beta^K{}_K)^2&=0 \label{scal-f}
\end{align}
---clearly, we obtained these equations by inserting $\beta_I=0$ into the vector \eqref{vec} and scalar \eqref{scal} constraints and by rewriting the rotation constraint unchanged. Conversely, if the fields $(\theta^I,r_J)$ satisfy the equations above, then the corresponding metric $q$ (now $q$ does not have to be real-analytic) and curvature $K$ satisfy the vacuum Einstein constraints. Therefore the system of equations \eqref{vec-f} and \eqref{scal-f} can be treated as a first order form of the Einstein constraints, at least in the case of real-analytic metrics. 

There is a partial symmetry in the equations \eqref{vec-f} and \eqref{scal-f} regarding the two-forms $(d\theta^I)$ and $(r_J)$. Taking into account that the condition $\beta^{I}=0$ is equivalent to $\beta^{[IJ]}=0$ we see that the equations except the vector constraint are invariant with respect to the exchange $(d\theta^I)\leftrightarrow (r^I)$. 

Let us also note that the l.h.s. of the scalar constraint \eqref{scal-f} can be seen as a sum of two values of the same quadratic form of signature $(5,1,0)$ defined on $3\times3$ symmetric matrices. It follows from Equation \eqref{R-sc} that now the value of this form on the matrix $(\beta^{IJ})=(\beta^{(IJ)})$ is (up to a factor) the Ricci scalar of the metric $q$:   
\begin{equation}
\rho=-\frac{1}{4}\Big(\ovcirc{\beta}^{IJ}\ovcirc{\beta}_{IJ}-\frac{1}{6}(\beta^K{}_K)^2\Big).
\label{Ricci-qdr}
\end{equation}

There is another algebraic interpretation of the r.h.s. of the equation above which is worth mentioning. The components $(\Gamma_{IJK})$ of the Levi-Civita connections one-forms $(\Gamma_{IJ})$ in the coframe $(\theta^K)$ are antisymmetric in their first two indices (see the second Equation in \eqref{Gamma}). Therefore the connection one-forms admit a pointwise decomposition into irreducible representation of $SO(3)$ analogous to that of $(d\theta^I)$:    
\[
\Gamma_{IJK}=\frac{1}{2}\eps_{IJ}{}^L\gamma_{LK}=\frac{1}{2}\eps_{IJ}{}^L \Big(\gamma_{[LK]}+\ovcirc{\gamma}_{LK}+\frac{1}{3}\gamma^N{}_N \delta_{LK}\Big),
\]
where $\ovcirc{\gamma}_{LK}:=\gamma_{(LK)}-\gamma^N{}_N \delta_{LK}/3$. It turns out that the irreducible parts of the connection one-forms are proportional (with different factors) to the corresponding parts of the derivatives $(d\theta^I)$---see Equations \eqref{ggg-bbb} for details. This means in particular that the condition $\beta^{[IJ]}=0$ implies that the matrix $(\gamma_{IJ})$ is symmetric and diagonalizable thereby. As noted in \cite{bry}, the Ricci scalar $\rho$ can be expressed by means of the second elementary function of the symmetric matrix $(\gamma_{IJ})$. Indeed, if $\la_1$, $\la_2$ and $\la_3$ are the eigenvalues of $(\gamma_{IJ})$, then Equation \eqref{Ricci-qdr} can be transformed by means of \eqref{ggg-bbb} to\footnote{In \cite{bry} instead of $(\gamma_{IJ})$ the matrix $(\gamma_{IJ}/2)$ is used (and denoted by $(S_{kl})$). This gives slightly different relation between the Ricci scalar and the second elementary function.}  
\[
\rho=\frac{1}{2}(\la_1\la_2+\la_2\la_3+\la_3\la_1).
\]

%***************************************************
\subsection{Dual frames}
%***************************************************

Let $(\theta^I)$ be a (local) coframe on $\Sigma$. We know that the antisymmetric part of $(d\theta^I)$ is zero, $\beta_I=0$, if and only if the dual frame $(\veps_J)$ satisfies the second equation in \eqref{eIE}. It turns out that it is not difficult to derive a general local form of such a frame. 

To this end let us express the tensor field $\Ec$ in a (local) coordinate system $(x^i)$:  
\[
\Ec=E(x^j)\,\partial_{x^1}\we\partial_{x^2}\we\partial_{x^3},
\]
where $E$ is a function. If $X=X^i\partial_{x^i}$, then  
\[
\Lc_X\Ec=(E_{,i}X^i-EX^i{}_{,i})\,\partial_{x^1}\we\partial_{x^2}\we\partial_{x^3}.
\]
The condition $\Lc_X\Ec=0$ is thus equivalent to the following equation: 
\begin{equation}
(\ln|E|)_{,i}X^i=X^i{}_{,i}.
\label{lnF}
\end{equation}

Since the vector field $\veps_3$ is non-zero everywhere, it is always possible to construct a local coordinate system $(x,y,z)$ such that   
\begin{align}
\veps_1&=a\partial_x+b\partial_y+f\partial_z, &\veps_2&=c\partial_x+e\partial_y+h\partial_z,& \veps_3&=\partial_z. 
\label{sym-fr}
\end{align}
Then 
\[
\Ec=\veps_1\we\veps_2\we\veps_3=E\,\partial_x\we\partial_y\we\partial_z
\]
with
\[
E=ae-bc.
\]

Now, applying Equation \eqref{lnF} we see that the three conditions $\Lc_{\veps_I}\Ec=0$ are equivalent to the following equations: 
\begin{align}
a(\ln|E|)_{,x}+b(\ln|E|)_{,y}+f(\ln|E|)_{,z}&=a_{,x}+b_{,y}+f_{,z},\label{lnF-f}\\
c(\ln|E|)_{,x}+e(\ln|E|)_{,y}+h(\ln|E|)_{,z}&=c_{,x}+e_{,y}+h_{,z},\label{lnF-h}\\
(\ln|E|)_{,z}&=0.\nonumber
\end{align}
Note that the last equation means that \emi $E$ is a function of $x$ and $y$ only:
\begin{equation}
ae-bc=E(x,y)\neq 0,
\label{F}
\end{equation}
and \emii the terms with $f$ and $h$ at the l.h.s. of Equations \eqref{lnF-f} and \eqref{lnF-h} disappear. Since $E$ does not depend on either $f$ or $h$, these two equations can be solved for $f$ and $h$ as follows:
\begin{align}
f&=\int\big(a(\ln|E|)_{,x}+b(\ln|E|)_{,y}-a_{,x}-b_{,y}\big)dz+\tilde{f}(x,y)=\label{f-int}\\
&= (\ln|E|)_{,x}\int a\,dz + (\ln|E|)_{,y}\int b\,dz -\int(a_{,x}+b_{,y})dz+\tilde{f}(x,y),\nonumber\\
h&=\int\big(c(\ln|E|)_{,x}+e(\ln|E|)_{,y}-c_{,x}-e_{,y}\big)dz+\tilde{h}(x,y)=\label{h-int}\\
&= (\ln|E|)_{,x}\int c\,dz + (\ln|E|)_{,y}\int e\,dz -\int(c_{,x}+e_{,y})dz+\tilde{h}(x,y).\nonumber
\end{align}

We thus conclude that every frame $(\veps_I)$ satisfying $\Lc_{\veps_I}\Ec=0$ can be always expressed locally in the form \eqref{sym-fr}, where the functions $a$, $b$, $c$ and $e$ are restricted by the condition \eqref{F}, while the functions $f$ and $h$ are given by the integrals \eqref{f-int} and \eqref{h-int} respectively.       

An other conclusion is that given a non-zero vector field $\veps_3$, there are many vector fields $\veps_1$ and $\veps_2$, which altogether satisfy $\Lc_{\veps_I}\Ec=0$---the set of all such vector fields $\veps_1$ and $\veps_2$ is labelled by six functions $a$, $b$, $c$, $e$, $\tilde{f}$ and $\tilde{h}$ such that
\begin{align*}
&ae-cd\neq 0, &&\veps_3(ae-cd)=\veps_3\tilde{f}=\veps_3\tilde{h}=0.
\end{align*}

Let us note finally that if the frame $(\veps_I)$ derived above is orthonormal with respect to a Riemannian metric $q$, then by virtue of the second equation in \eqref{div-q}, each vector field $\veps_I$ is divergence-free.

%***************************************************
\section{Other gauges}
%***************************************************

The Einstein constraints in the form \eqref{pri-con}-\eqref{scal} are imposed not on a Riemannian metric on $\Sigma$, but rather on its orthonormal coframe. This gives a possibility to describe the same initial data in terms of different orthonormal coframes, which means that there is a local $SO(3)$ gauge freedom here. In the previous section we used this freedom to zero the second order terms in the scalar constraint. Below we will describe briefly two other gauges found in the literature.

%***************************************************
\subsection{Darboux gauge}
%***************************************************

It was proven by Darboux \cite{darboux} (see also \cite{metr-diag}) that for every Riemannian metric $q$ on $\Sigma$ it is possible to find around every point of the manifold, a local coordinate system $(x,y,z)$, such that the metric is diagonal in the coordinates:  
\begin{equation}
q=A^2dx^2+B^2dy^2+C^2dz^2
\label{q-diag}
\end{equation}
for some everywhere non-zero functions $A$, $B$ and $C$. This implies the existence of a local orthonormal coframe of $q$ of the following simple form:     
\begin{align}
\theta^1&=A\,dx, & \theta^2&=B\,dy, & \theta^3&=C\,dz.
\label{darb-cfr}
\end{align}
It follows from these equations that
\begin{equation}
d\theta^I\we\theta^I=0
\label{dtI-tI}
\end{equation}
for every $I$. This result together with Equation \eqref{ttIJ-bIJ} allows us to conclude that for the coframe \eqref{darb-cfr} all the diagonal components of $(\beta^{IJ})$ vanish: 
\begin{equation}
\beta^{II}=0
\label{bII=0}
\end{equation}
---let us emphasize that there is no summation over $I$ in the equation above.

On the other hand, let us suppose that \eqref{bII=0} holds for $I=1$. This by virtue of Equation \eqref{ttIJ-bIJ} gives us \eqref{dtI-tI} with $I=1$. Let us consider two cases: $d\theta^1=0$ and $d\theta^1\neq 0$. In the first case, by virtue of the Poincare lemma we have locally  $\theta^1=dx$ for a function $x$. In the second case we note that $d\theta^1\we d\theta^1=0$ since $\Sigma$ is $3$-dimensional. This means that $d\theta^1$ is of rank\footnote{Let $\omega$ be a two-form. We say that $\omega$ is of rank $r$, if \emi the exterior product of $r$ factors $\omega$ is non-zero and \emii the exterior product of $r+1$ factors $\omega$ is zero.} $1$. Then $d\theta^1=A\,dx$ for some functions $A$ and $x$---this form of $d\theta^1$ is implied by the following theorem
\cite{exact-sol}:
\begin{thr}
Suppose $\alpha$ is a one-form on an $m$-dimensional manifold such that its derivative $d\alpha$ is of rank $r$ and the product of $\alpha$ and $r$ factors $d\alpha$ is zero: $\alpha\we d\alpha\we\ldots\we d\alpha=0$. Then there exists a local coordinate system $(y^1,y^2,\ldots,y^r,x^1,\ldots,x^{m-r})$ such that
\[
\alpha=y^1\,dx^1+\ldots+y^r\,dx^r.
\]  
\end{thr}
\noindent We thus see that if $\beta^{11}=0$, then in general $d\theta^1=A\,dx$.

Consequently, if \eqref{bII=0} holds for every $I$, then \eqref{darb-cfr} holds on a common domain of all the functions $A$, $B$, $C$ and $x$, $y$, $z$. Since $(\theta^I)$ is a coframe, then $dx\we dy\we dz\neq 0$, which means that the functions $(x,y,z)$ form a local coordinate system.     

We conclude that the condition \eqref{bII=0} is equivalent to \eqref{darb-cfr} and that it implies the diagonal form \eqref{q-diag} of the metric $q$. 

In \cite{tegr-con} a general local form of a coframe was found that satisfies both gauges $\beta_I=0$ and \eqref{bII=0}. Each coframe of this sort is described by three functions $F$, $G$ and $H$ of coordinates $(x,y,z)$ such that each function depends merely on two of the three coordinates: 
\begin{align*}
\theta^1&=\pm\frac{G(y,z)}{F(x,y)H(x,z)}\,dx, & \theta^2&=\pm\frac{H(x,z)}{G(y,z)F(x,y)}\,dy, & \theta^3&=\pm\frac{F(x,y)}{H(x,z)G(y,z)}\,dz.
%\label{ABC}
\end{align*}

%***************************************************
\subsection{Nester gauge}
%***************************************************

Both gauges considered so far, $\beta_I=0$ and $\beta^{II}=0$, impose first order PDE's on a coframe. Nester in \cite{nester-on-1,nester-on-2,nester-on-3} considered a gauge defined by some second order PDE's imposed on a coframe $(\theta^I)$:  
\begin{align}
d\tilde{Q}&=0, & *\,d*\hat{Q}&=0,
\label{nest-gg}
\end{align}
where
\begin{align*}
\tilde{Q}&:=\veps_I\lr d\theta^I, & \hat{Q}&:=\theta_I\we d\theta^I.
\end{align*}
Using \eqref{dth} and the first equation in \eqref{bI-bJIJ} one obtains
\begin{align*}
\tilde{Q}&=-\beta_I\,\theta^I, & \hat{Q}&=\frac{1}{2}\beta_{IJK}\,\theta^I\we\theta^J\we\theta^K.
\end{align*}
Let us now insert these two formulas to \eqref{nest-gg}. Applying Equations \eqref{b-dec}--\eqref{*r*th}, \eqref{var-eps} and
\[
*(\theta^I\we\theta^J)=\eps^{IJK}\theta_K
\]
we rewrite the conditions \eqref{nest-gg} in terms of the irreducible parts of $(d\theta^I)$:  
\begin{align*}
\eps^{IJK}\beta_{K,J}+\frac{1}{2}\ovcirc{\beta}^{IJ}\beta_J+\frac{1}{6}\beta^K{}_K\beta^I&=0, & \beta^K{}_{K,I}&=0.
\end{align*}

The above form of the conditions show that the Nester gauge is essentially different from both first order gauges considered above. It seems also that in general the Nester gauge does not annihilate the second order term $\beta^I{}_{,I}$ in the scalar constraint \eqref{scal}.

It was proven in \cite{nester-dirac} that, given a Riemannian metric on $\Sigma$, the existence and uniqueness of its orthonormal coframe satisfying the Nester conditions \eqref{nest-gg} is equivalent to the existence and uniqueness of solutions to a Dirac equation.

%***************************************************
\section{Summary and outlook}
%***************************************************

In this paper we showed that the TEGR constraints in the gauge $\xi^I=0$ considered in \cite{tegr-con} are equivalent to the Einstein constraints in vacuum with zero cosmological constant. The transition from the former constraints to the latter ones was achieved by replacing the one-forms (the coframe) $(\theta^I)$ and the conjugate momenta $(r_J)$ by the Riemannian metric $q$ and the extrinsic curvature $K$ according to the following formulas:   
\begin{align*}
  q&=\delta_{IJ}\,\theta^I\ot\theta^J=\theta_I\ot\theta^I, & K&=\theta^I\ot*r_I-\frac{1}{2}*(r^K\we\theta_K)\,\theta_I\ot\theta^I.
\end{align*}
The last formula is equivalent to the previously used equations \eqref{r-K} and it shows explicitly that $K$ is actually a fairly non-trivial function of both variables $(\theta^I)$ and $(r_J)$.     

The TEGR constraints can be thus regarded as a  ``teleparallel'' form of the Einstein constraints. It seems that this form have not yet been subjected to deeper analysis.

The scalar constraint in both standard and ``teleparallel'' form is a PDE of second order. The order of the ``teleparallel'' scalar constraint can be reduced by $1$ at the cost of imposing first order equations $\beta_I=0$ on the coframe \cite{tegr-con}. We invoked here the Bryant theorem \ref{bry-thr} to prove that these equations can be always solved locally, if the spatial metric $q$ is real-analytic. This led us to the conclusion that the standard Einstein constraints imposed on a real-analytic metric are locally equivalent to the system \eqref{vec-f}--\eqref{scal-f} of first order PDE's. This system certainly deserves further study. In particular, it would be interesting to check whether the equivalence holds also for smooth metrics (or even for ones of regularity lower than smooth). Another intriguing property of the system \eqref{vec-f}--\eqref{scal-f} worth of closer look, is its partial symmetry w.r.t. to the exchange $(d\theta^I)\leftrightarrow (r^I)$.  

We also found a general form of a frame such that its dual coframe satisfy the equations $\beta_I=0$. 

In the forthcoming paper \cite{prep-ao-js} we will apply the ``teleparallel'' form of the Einstein constraints and the system \eqref{vec-f}--\eqref{scal-f} to derive some exact solutions to the constraints.

\paragraph{Acknowledgments} We are very grateful to Piotr Chru\'sciel, Pawe{\l} Nurowski and Adam Szereszewski for valuable discussions, hints and help.

%***************************************************
\appendix
%***************************************************

%***************************************************
\section{The momenta $(r_I)$ versus the extrinsic curvature \label{r-K-app}}
%***************************************************

In this section we will derive a relation between the momenta $(r_I)$ and the extrinsic curvature of the manifold $\Sigma$ regarded as a spatial slice of a spacetime. The relation will be obtained based on the gauge $\xi^I=0$.    

In \cite{oko-tegr-I} a Hamiltonian formulation of TEGR was derived, where a point in the TEGR phase space was described as a collection of four one-forms $(\theta^A)$ and four two-forms $(p_B)$, $A,B\in\{0,1,2,3\}$, defined on $\Sigma$. Then in \cite{ham-nv} the variables $(\theta^A,p_B)$ were transformed into the variables $(\xi^I,\theta^J,r_K,\zeta_L)$ used in the present paper. Below to achieve our goal we will refer to both the Hamiltonian formulation of TEGR presented in \cite{oko-tegr-I} and to the canonical transformation introduced in \cite{ham-nv}. 

The departure point of \cite{oko-tegr-I} is a Lagrangian formulation of TEGR, where the configuration variables are four one-forms $(\bld{\theta}^A)$ constituting a global coframe on a (four dimensional) manifold $\Mc\cong\R\times\Sigma$. The forms are assumed to define a spacetime (Lorentzian) metric
\begin{equation}
g=\eta_{AB}\,\bld{\theta}^A\ot\bld{\theta}^B, \quad (\eta_{AB})={\rm diag}(-1,1,1,1),
\label{g-metr}
\end{equation}
such that the metric $q$ induced on $\{t\}\times\Sigma\subset \Mc$ is spatial (Riemannian) for every moment $t\in\R$. In the formalism the matrix $(\eta_{AB})$ and its inverse $(\eta^{AB})$ are used to, respectively, lower and raise up indices $A,B,C,D\in\{0,1,2,3\}$.   

Once $3+1$ decomposition of the fields $(\bld{\theta}^A)$ w.r.t. the product $\R\times\Sigma$ is done, we are left with $t$-dependent fields $(N,\vec{N},\theta^A)$ on $\Sigma$, where $N$ is the (standard) lapse function, $\vec{N}$ the (standard) shift vector field, while $(\theta^A)$ are one-forms on $\Sigma$. Then a suitable Legendre transformation gives the following momentum conjugate to the form $\theta^A$:         
\begin{multline}
p_A=N^{-1}\Big({\theta}_B\we{*}[{\dot{\theta}}^B\we{\theta}_A-N({d}\bld{\xi}^B\we{\theta}_A-{d}{\theta}^B\we\bld{\xi}_A)-\Lc_{\vec{N}}{\theta}^B\we{\theta}_A ]-\\-\frac{1}{2}{\theta}_A\we{*}[{\dot{\theta}}^B\we{\theta}_B-N({d}\bld{\xi}^B\we{\theta}_B-{d}{\theta}^B\we\bld{\xi}_B)-\Lc_{\vec{N}}{\theta}^B\we{\theta}_B]\Big),
\label{p_A-exp}
\end{multline}
being a two-form on $\Sigma$. In the formula above, $\dot{\theta}^A$ denotes the time derivative of $\theta^A$ (i.e. the derivative of $\theta^A$ w.r.t. $t$), $\Lc_{\vec{N}}$ does the Lie derivative on $\Sigma$ w.r.t. $\vec{N}$, and $\bld{\xi}^A$ is a function on $\Sigma$ defined as  
\begin{equation}
\bld{\xi}^A:=-\frac{1}{3!}\veps^A{}_{BCD}*(\theta^B\we\theta^C\we\theta^D),
\label{bxi}
\end{equation}
where \emi $\veps_{ABCD}$ is a totally antisymmetric symbol such that $\veps_{0123}=1$, and \emii $*$ denotes the Hodge ``star'' given by the spatial metric $q$.     

In \cite{ham-nv} new canonical variables $(\xi^I,\theta^J,r_K,\zeta_L)$ were introduced---they are related to the original variables $(\theta^A,p_B)$ as follows: 
\begin{align}
\theta^0&=\sgn(\theta^J)\frac{\xi_I}{\sqrt{1+\xi_K\xi^K}}\,\theta^I, & \theta^I&=\theta^I,\label{th0}\\
p_0&=\sgn(\theta^J)\sqrt{1+\xi_K\xi^K}\vec{\theta}^I\lr\zeta_I, & p_I&=r_I-\xi_I\,\vec{\theta}^J\lr\zeta_J,\label{p_0}
\end{align}
where 
\[
\sgn(\theta^I):=
\begin{cases}
1 & \text{if $(\theta^I)$ is compatible with the orientation of $\Sigma$}\\
-1 & \text{otherwise}
\end{cases}.
\]
The spatial metric $q$, when expressed in terms of the new variables, takes the following form: 
\begin{equation}
q=q_{IJ}\,\theta^I\ot\theta^J=\Big(\delta_{IJ}-\frac{\xi_I\xi_J}{1+\xi_K\xi^K}\Big)\theta^I\ot\theta^J.
\label{q-xi}
\end{equation}

Suppose now that the original variables $N$, $\vec{N}$, $(\theta^I)$ and $(\dot{\theta}^J)$ give raise to canonical variables $(\xi^I=0,\theta^J,r_K,\zeta_L)$. Then by virtue of \eqref{th0}, \eqref{p_0} and \eqref{bxi}    
\begin{align*}
r_I&=p_I, & \theta^0&=0, & \bld{\xi}^I&=-\frac{1}{2}\veps^I{}_{0JK}*(\theta^0\we\theta^J\we\theta^K)=0. 
\end{align*}
Inserting these three equations to \eqref{p_A-exp} gives us
\begin{equation}
r_I={\theta}_J\we{*}\Big[\frac{{\dot{\theta}}^J-\Lc_{\vec{N}}{\theta}^J}{N}\we{\theta}_I\Big]-\frac{1}{2}{\theta}_I\we{*}\Big[\frac{{\dot{\theta}}^J-\Lc_{\vec{N}}{\theta}^J}{N}\we{\theta}_J\Big].
\label{r_I-exp}
\end{equation}
Denoting
\[
\frac{{\dot{\theta}}^J-\Lc_{\vec{N}}{\theta}^J}{N}\equiv\alpha^J=\alpha^J{}_I\,\theta^I
\]
we transform \eqref{r_I-exp} further\footnote{We assumed that $\xi^I=0$. Therefore now by virtue of \eqref{q-xi} the coframe $(\theta^J)$ is orthonormal w.r.t. the metric $q$.}:
\[
r_I=\alpha_{JK}\Big(\eps^{K}{}_{IL}\,\theta^J\we\theta^L-\frac{1}{2}\eps^{KJ}{}_L\,\theta_I\we\theta^L\Big).
\]
Applying the ``star'' $*$ to both sides of the equation above we obtain 
\begin{multline*}
*r_I=\alpha_{JK}\Big(\eps^{K}{}_{IL}\eps^{JL}{}_N-\frac{1}{2}\eps^{KJ}{}_L\eps_I{}^L{}_N\Big)\theta^N=\alpha_{JK}(-\delta^{KJ}\delta_{IN}+\delta^K{}_N\delta^J{}_I-\delta^{K}{}_{[N}\delta^J{}_{I]})\theta^N=\\=(-\alpha^K{}_K\delta_{IN}+\alpha_{(IN)})\theta^N.
\end{multline*}
Comparing this result with first equation in \eqref{*r*th} we conclude that if $\xi^I=0$, then 
\begin{equation}
r_{IN}=2(\alpha_{(IN)}-\alpha^K{}_K\delta_{IN}).
\label{r-al}
\end{equation}

Note that we got symmetric components: $r_{IJ}=r_{JI}$, which means that the corresponding momenta $(r_I)$  satisfy the rotation constraint. This is not a surprise because the momenta are obtained via the Legendre transformation and therefore they must satisfy that constraint being a primary one.

On the other hand, the $3+1$ decomposition of the spacetime metric \eqref{g-metr} w.r.t. to the product $\R\times\Sigma$ gives us the same lapse function $N$, the same shift vector field $N$ and the Riemannian metric $q$ as $t$-dependent fields on $\Sigma$. It is well known that the extrinsic curvature $K$ of $\Sigma$ embedded in the spacetime $\Mc$ can be expressed as        
\begin{equation}
K=\frac{\dot{q}-\Lc_{\vec{N}}q}{2N},
\label{K-curv}
\end{equation}
where again the dot over $q$ denotes the derivative w.r.t. time $t$.   

To calculate $\dot{q}$ we use \eqref{q-xi} and obtain 
\begin{multline*}
\dot{q}=\frac{(\dot{\xi}_I\xi_J+\xi_I\dot{\xi}_J)(1+\xi^K\xi_K)-2\,\xi_I\xi_J\dot{\xi^K}\xi_K}{(1+\xi^K\xi_K)^2}\,\theta^I\ot\theta^J+\\+\Big(\delta_{IJ}-\frac{\xi_I\xi_J}{1+\xi^K\xi_K}\Big)(\dot{\theta}^I\ot\theta^J+\theta^I\ot\dot{\theta^J}).
\end{multline*}
But if $\xi^I=0$, then the above formula reduces to 
\[
\dot{q}=\dot{\theta}^I\ot\theta_I+\theta_I\ot\dot{\theta^I}.
\]
Similarly, if $\xi^I=0$, then 
\[
\Lc_{\vec{N}}q=(\Lc_{\vec{N}}{\theta}^I)\ot\theta_I+\theta_I\ot\Lc_{\vec{N}}{\theta^I}.
\]

Inserting the last two results to \eqref{K-curv} we obtain
\[
K=\frac{{\dot{\theta}}^I-\Lc_{\vec{N}}{\theta}^I}{2N}\ot \theta_I+\theta_I\ot\frac{{\dot{\theta}}^I-\Lc_{\vec{N}}{\theta}^I}{2N}=\frac{1}{2}(\alpha^I\ot\theta_I+\theta_I\ot\alpha^I)=\alpha_{(IN)}\theta^I\ot\theta^N.
\]
Thus in the gauge $\xi^I=0$ the components of $K$ in the coframe $(\theta^I)$ read   
\[
K_{IN}=\alpha_{(IN)}.
\]
Inserting this into \eqref{r-al} we obtain
\[
r_{IJ}=2(K_{IJ}-K^L{}_L\delta_{IJ}).
\]
Consequently,
\begin{align*}
r^L{}_L&=-4 K^L{}_L, & \ovcirc{r}_{IJ}&=2\ovcirc{K}_{IJ},\\
r_{IJ}&=2\ovcirc{K}_{IJ}-\frac{4}{3}K^L{}_L\delta_{IJ}, & K_{IJ}&=\frac{1}{2}r_{IJ}-\frac{1}{4}r^K{}_K\delta_{IJ}. 
\end{align*}
where $\ovcirc{K}_{IJ}:=K_{IJ}-K^L{}_L\,\delta_{IJ}/3$ is the traceless part of $K$.  

%***************************************************
\section{Ricci tensor and Ricci scalar \label{ricci-app}}
%***************************************************

Suppose that $(\theta^I)$ is a (local) orthonormal coframe of a Riemannian metric $q$ on $\Sigma$. Below we will express the components $(\rho_{IJ})$ of the Ricci tensor and the Ricci scalar $\rho$ of the metric as functions of the irreducible parts of the derivatives $(d\theta^I)$.      

Denote by $(\rho_{IJKL})$ the components of the (covariant) Riemann tensor of $q$ in the coframe $(\theta^I)$. As it is well-known
\[
\frac{1}{2}\rho_{IJKL}\,\theta^K\we\theta^L=d\,\Gamma_{IJ}+\Gamma_{IM}\we\Gamma^M{}_J,
\]
where
\[
\Gamma_{IJ}=\Gamma_{IJK}\,\theta^K
\]
are the Levi-Civita connection one-forms of $q$ expressed in the coframe $(\theta^I)$. Inserting the last equation into the previous one give us 
\[
\rho_{IJKL}=\Gamma_{IJL,K}-\Gamma_{IJK,L}+\Gamma_{IJM}\beta^M{}_{KL}+\Gamma_{IMK}\Gamma^M{}_{JL}-\Gamma_{IML}\Gamma^M{}_{JK}.
\]
Therefore the components of the Ricci tensor in the coframe $(\theta^I)$ read
\begin{equation}
\rho_{JL}=\rho^I{}_{JIL}=\Gamma^I{}_{JL,I}-\Gamma^I{}_{JI,L}+\Gamma^I{}_{JM}\beta^M{}_{IL}+\Gamma^I{}_{MI}\Gamma^M{}_{JL}-\Gamma^I{}_{ML}\Gamma^M{}_{JI}.
\label{ricci-t}
\end{equation}

In order to express the components $(\rho_{JL})$ as functions of the irreducible parts of $(d\theta^I)$ let us first describe the connection coefficients $(\Gamma_{IJK})$ in terms of these parts. Inserting the second equation in \eqref{dth} into the first equation in \eqref{Gamma} we obtain
\[
\frac{1}{2}\beta_{IJK}\theta^J\we\theta^K+\Gamma_{IJK}\theta^K\we\theta^J=0,
\]
which implies that
\begin{equation}
\beta_{IJK}=\Gamma_{IJK}-\Gamma_{IKJ}.
\label{b-GG}
\end{equation}
By virtue of the second equation in \eqref{Gamma}, we have the following decomposition:
\begin{align}
\Gamma_{IJK}&=\frac{1}{2}\eps_{IJ}{}^L\gamma_{LK}, & \gamma_{LK}&=\gamma_{[LK]}+\ovcirc{\gamma}_{LK}+\frac{1}{3}\gamma^N{}_N \delta_{LK}=\eps_L{}^{IJ}\Gamma_{IJK},
\label{Gamma-dec}
\end{align}
where $(\ovcirc{\gamma}^L{}_{K})$ is a field on $\Sigma$ valued in traceless symmetric $3\times 3$ matrices. Contracting both sides of \eqref{b-GG} with $\eps^{JK}{}_L$ and applying the first equation in \eqref{b-dec} and the decomposition \eqref{Gamma-dec} we obtain  
\begin{equation*}
\beta_{IL}=2\Gamma_{IJK}\eps^{JK}{}_L=\eps_{IJ}{}^N\gamma_{NK}\eps^{JK}{}_L=-\gamma_{LI}+\gamma^N{}_N\delta_{IL}=\gamma_{[IL]}-\ovcirc{\gamma}_{IL}+\frac{2}{3}\gamma^N{}_N \delta_{IL}.
\end{equation*}
Comparing this result with the second equation in \eqref{b-dec} we conclude that
\begin{align}
\gamma_{[IL]}&=\beta_{[IL]}, & \ovcirc{\gamma}_{IL}&=-\ovcirc{\beta}_{IL}, & \gamma^N{}_N&=\frac{1}{2}\beta^N{}_N.
\label{ggg-bbb}
\end{align}
Consequently,
\[
\gamma_{IL}=\beta_{[IL]}-\ovcirc{\beta}_{IL}+\frac{1}{6}\beta^N{}_N\delta_{IL}=\eps_{ILN}\beta^N-\ovcirc{\beta}_{IL}+\frac{1}{6}\beta^N{}_N\delta_{IL}
\]
and
\begin{equation}
\Gamma_{IJK}=\frac{1}{2}\Big(\delta_{IK}\beta_J-\delta_{JK}\beta_I-\eps_{IJ}{}^M\ovcirc{\beta}_{MK}+\frac{1}{6}\eps_{IJK}\beta^N{}_N\Big).
\label{G-beta-dec}
\end{equation}

Inserting the decompositions \eqref{G-beta-dec} and \eqref{b-dec} to \eqref{ricci-t} and taking into account that the Ricci tensor is symmetric, $\rho_{JL}=\rho_{(JL)}$, one obtains (after tedious calculations) 
\begin{multline*}
\rho_{JL}=-\frac{1}{2}\Big(\beta_{(J,L)}+\delta_{JL}\beta^K{}_{,K}+\eps^{IK}{}_{(J}\ovcirc{\beta}_{L)I,K}+\delta_{JL}\beta^K\beta_K+\eps^{IK}{}_{(J}\ovcirc{\beta}_{L)I}\beta_K-\ovcirc{\beta}_{JK}\ovcirc{\beta}^{K}{}_L+\\+\frac{1}{2}\delta_{JL}\ovcirc{\beta}_{IK}\ovcirc{\beta}^{IK}-\frac{1}{6}\ovcirc{\beta}_{JL}\beta^K{}_K-\frac{1}{36}\delta_{JL}(\beta^K{}_K)^2\Big).
\end{multline*} 
This formula implies the following form of the Ricci scalar:
\[
\rho=\rho_{JL}\delta^{JL}=-\frac{1}{4}\Big(8\beta^K{}_{,K}+6\beta^K\beta_K+\ovcirc{\beta}_{IK}\ovcirc{\beta}^{IK}-\frac{1}{6}(\beta^K{}_K)^2\Big).
\]

%***************************************************
%***************************************************
%\bibliography{bibliography}{}
%\bibliographystyle{oko}
%***************************************************
%***************************************************

\end{document}